\RequirePackage{fix-cm}
\documentclass[twocolumn]{svjour3}          
\smartqed  
\usepackage{graphicx,amsmath}
\usepackage{bm}       
\usepackage{graphicx,color,tikz} 
\usepackage{hyperref} 
\begin{document}

\title{Thermodynamic properties of disordered quantum spin ladders}


\author{Erol Vatansever         \and Georgi G. Grahovski \and Nikolaos G. Fytas}


\institute{Erol Vatansever   \at
              {\it Department of Physics, Dokuz Eyl\"{u}l University, Izmir (Turkey)} \\
              \email{erol.vatansever@deu.edu.tr}           
           \and
           Georgi G. Grahovski \at
              {\it School of Mathematics, Statistics and Actuarial Science, University of Essex, Colchester (UK)} \\
              \email{grah@essex.ac.uk} 
               \and
           Nikolaos G. Fytas \at
              {\it School of Mathematics, Statistics and Actuarial Science, University of Essex, Colchester (UK)} \\
              \email{nikolaos.fytas@essex.ac.uk} 
}

\date{Received: date / Accepted: date}

\maketitle

\begin{abstract}
In this paper, we study the thermodynamic properties of spin-$1/2$ antiferromagnetic Hei\-sen\-berg ladders  by means of the stochastic series expansion quantum Monte Carlo technique. This includes the thermal properties of the specific heat, uniform and staggered susceptibilities, spin gap, and structure factor. Our numerical simulations are probed over a large ensemble of random realizations in a wide range of disorder strengths $r$, from the clean ($r=0$) case up to the diluted ($r \rightarrow 1$) limit, and for selected choices of number of legs $L_y$ per site. Our results show some interesting phenomena, like the presence of crossing points in the temperature plane for both the specific heat and uniform susceptibility curves which appear to be universal in $r$, as well as a variable dependence of the spin gap in the amount of disorder upon increasing $L_y$.
\end{abstract}

\section{Introduction}
\label{sec:intro}

Understanding the effect of quenched randomness on zero- and finite-temperature properties of spin models is one of the most intriguing problems in theoretical physics. Recent model systems including quenched randomness that are under investigation include the spin-$1/2$ Heisenberg spin chains \cite{shu16,shiroka19}, the spin-1/2 $J-Q$ model on the
square lattice \cite{liu18}, and quantum spin chains with power-law long-range antiferromagnetic couplings \cite{moure18}. Low-dimensional spin systems have also been at the centre of intense investigation due to the
development of advanced theoretical, computational and experimental methods that paved the way for new results \cite{suzuki15,patel16,zhang17,akimitsu19,yu20}. Among the wealth of low-dimensional systems, quantum spin systems with antiferromagnetic interactions share plentiful physical properties, even in one dimension. 
For example, Haldane's conjecture that antiferromagnetic spin chains with integer spin values exhibit a gapped spectrum 
has been backed up by solid theoretical \cite{affleck86}, numerical \cite{nightingale86}, and experimental \cite{buyers86} studies. 
It is worth noting here that the spin-$1/2$ Heisenberg coupled chains with an even number of legs have a finite spin gap ($\Delta$) to the lowest triplet excitation. Also, some ladder systems have an exponentially decaying spin-spin correlation function and uniform 
susceptibility, manifesting the existence of a spin gap \cite{frischmuth96,sandvik10}. The width of the spin gap for a two-leg ladder spin system with the isotropic coupling constant $J$ can be roughly estimated as $\Delta \sim 0.5J$ via quantum Monte Carlo techniques \cite{barnes93}. What is more, the value of the spin gap can be modified or even completely eliminated in various ways, among others, by the choice of a different lattice topology \cite{melin02,maiti18}, the diffusion of disorder in the spin-spin coupling  \cite{metavitsiadis17}, or the application of an external magnetic field \cite{chitra97}. 

Different ladder models have been studied in the literature, such as spin-ladder systems with dimerization \cite{cabra00,chen14,kariyado15,naseri17}, zig-zag ladders \cite{hoyos04,bunder09}, and mixed ladders \cite{aristov04,batchelor04,zad18}. The presence of quenched randomness is expected, in most cases, to have a significant effect on the thermal and magnetic properties of the system, even at the low disorder limit. Weakly disordered anisotropic spin-$1/2$ ladders have been handled perturbatively \cite{orignac98} and critical properties of 
strongly disordered systems have been analyzed benefiting from the strong disorder (density matrix) renormalization-group 
method \cite{melin02,igloi05,vojta13}. 

Modern quantum Monte Carlo techniques are powerful and robust tools for the study of disordered spin-ladder systems. For instance, the stochastic series expansion quantum Monte Carlo technique has already been successfully used to investigate the spin-$1/2$ Hei\-senberg ladders under the presence of quenched bond randomness \cite{trinh13}, indicating that neighbouring bond energies change sensitively with the position of disorder in the spin-spin coupling term. Furthermore, in Ref. \cite{hormann18}, some unusual and interesting effects of disorder on collective excitations have been reported with the calculation of the ground-state dynamic structure factor for a ladder system with bond disorder along the legs and rungs of the ladder. More recently, the thermodynamic properties of a two-leg quantum spin-ladder system including quenched bond randomness only along the rung direction of the system have been discussed \cite{Kanbur2}, for 
varying values of the system parameters. 

In this framework, we attempt to provide here an additional insight into some critical aspects of disordered quantum spin-ladder systems, not fully considered previously. Therefore, the goal of the present study is threefold: (i) Monitor the trend of the crossing points of the specific heat and uniform susceptibility curves, (ii) scan the response of the spin-gap parameter, and (iii) scrutinize the low-temperature behaviour of the staggered susceptibility and structure factor, while increasing the number of legs of the spin-ladder system and for different values of the disorder parameter. Whenever it is possible, we compare our results with the main outcomes of Ref. \cite{Kanbur2}. Before going on further, we would like to underline that the existence of the crossing point phenomenon, observed also in the present work, is one of the most interesting and 
puzzling effects in strongly correlated electron systems. Up to now, and to the best of our knowledge, no unified mechanism for 
this phenomenon has been proposed. However, it is generally believed that crossing points occur in certain systems that are in 
the vicinity of a quantum or a second-order phase transition, or some physical systems showing magnetic instability, 
such that their properties are prominently sensitive to changes in thermodynamic variables, such as the
magnetic field and the temperature.

The structure of the paper is  as follows: In Section \ref{sec:model} we briefly describe the model and the implementation of the stochastic series expansion quantum Monte Carlo method. In Section \ref{sec:numerical} we provide the simulation details and the physical observables studied here. The numerical results are presented and analysed in Section \ref{sec:results}. Finally, Section \ref{sec:summary} contains a summary and a brief outlook.

\section{Model and method}
\label{sec:model}

\subsection{Antiferromagnetic spin-1/2 random-bond Heisenberg spin ladders}
\label{ssec:model}

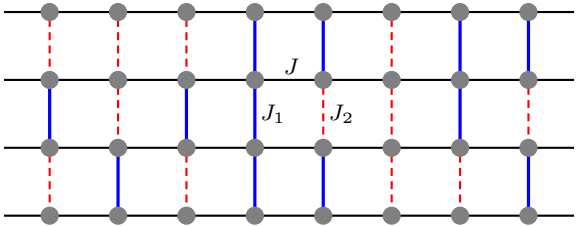
\begin{figure}%
\begin{center}
\begin{tikzpicture}[scale=1.2]

\draw[thick] (-0.5,0) -- (5.75,0);
\draw[thick] (-0.5,0.75) -- (5.75,0.75);
\draw[thick] (-0.5,1.5) -- (5.75,1.5);
\draw[thick] (-0.5,2.25) -- (5.75,2.25);

\draw[thick,red,densely dashed] (0,0) -- (0,0.75);
\draw[thick,red,densely dashed] (0,1.5) -- (0,2.25);
\draw[very thick,blue] (0,0.75) -- (0,1.5);

\draw[very thick,blue] (0.75,0) -- (0.75,0.75);
\draw[thick,red,densely dashed] (0.75,0.75) -- (0.75,1.5);
\draw[thick,red,densely dashed] (0.75,1.5) -- (0.75,2.25);

\draw[thick,red,densely dashed] (1.5,0) -- (1.5,0.75);
\draw[thick,red,densely dashed] (1.5,1.5) -- (1.5,2.25);
\draw[very thick,blue] (1.5,0.75) -- (1.5,1.5);

\draw[very thick,blue] (2.25,0) -- (2.25,0.75);
\draw[very thick,blue] (2.25,0.75) -- (2.25,1.5);
\draw[very thick,blue] (2.25,1.5) -- (2.25,2.25);

\draw[very thick,blue] (3,0) -- (3,0.75);
\draw[thick,red,densely dashed] (3,0.75) -- (3,1.5);
\draw[very thick,blue] (3,1.5) -- (3,2.25);

\draw[thick,red,densely dashed] (3.75,0) -- (3.75,0.75);
\draw[thick,red,densely dashed] (3.75,0.75) -- (3.75,1.5);
\draw[thick,red,densely dashed] (3.75,1.5) -- (3.75,2.25);

\draw[thick,red,densely dashed] (4.5,0) -- (4.5,0.75);
\draw[very thick,blue] (4.5,0.75) -- (4.5,1.5);
\draw[very thick,blue] (4.5,1.5) -- (4.5,2.25);

\draw[very thick,blue] (5.25,0) -- (5.25,0.75);
\draw[thick,red,densely dashed] (5.25,0.75) -- (5.25,1.5);
\draw[very thick,blue] (5.25,1.5) -- (5.25,2.25);


\draw (2.45,1.125) node {\small  $J_1$};
\draw (3.2,1.125) node {\small  $J_2$};
\draw (2.65,1.65) node {\small  $J$};

\fill[black!50] (0,0) circle(1.0mm);
\fill[black!50] (0.75,0) circle(1.0mm);
\fill[black!50] (1.5,0) circle(1.0mm);
\fill[black!50] (2.25,0) circle(1.0mm);
\fill[black!50] (3,0) circle(1.0mm);
\fill[black!50] (3.75,0) circle(1.0mm);
\fill[black!50] (4.5,0) circle(1.0mm);
\fill[black!50] (5.25,0) circle(1.0mm);

\fill[black!50] (0,0.75) circle(1.0mm);
\fill[black!50] (0.75,0.75) circle(1.0mm);
\fill[black!50] (1.5,0.75) circle(1.0mm);
\fill[black!50] (2.25,0.75) circle(1.0mm);
\fill[black!50] (3,0.75) circle(1.0mm);
\fill[black!50] (3.75,0.75) circle(1.0mm);
\fill[black!50] (4.5,0.75) circle(1.0mm);
\fill[black!50] (5.25,0.75) circle(1.0mm);

\fill[black!50] (0,1.5) circle(1.0mm);
\fill[black!50] (0.75,1.5) circle(1.0mm);
\fill[black!50] (1.5,1.5) circle(1.0mm);
\fill[black!50] (2.25,1.5) circle(1.0mm);
\fill[black!50] (3,1.5) circle(1.0mm);
\fill[black!50] (3.75,1.5) circle(1.0mm);
\fill[black!50] (4.5,1.5) circle(1.0mm);
\fill[black!50] (5.25,1.5) circle(1.0mm);

\fill[black!50] (0,2.25) circle(1.0mm);
\fill[black!50] (0.75,2.25) circle(1.0mm);
\fill[black!50] (1.5,2.25) circle(1.0mm);
\fill[black!50] (2.25,2.25) circle(1.0mm);
\fill[black!50] (3,2.25) circle(1.0mm);
\fill[black!50] (3.75,2.25) circle(1.0mm);
\fill[black!50] (4.5,2.25) circle(1.0mm);
\fill[black!50] (5.25,2.25) circle(1.0mm);

\end{tikzpicture}
\end{center}

\caption{An example of the quenched bond-disorder configuration on a quantum spin-ladder system. The couplings $J_1$ and $J_2$  along the rungs are depicted by solid blue and red dotted lines, respectively. The interactions $J$ are represented by solid black lines.}\label{fig1}
\end{figure}

We introduce the Hamiltonian of the quantum spin-ladder model in a general framework in order to align with the formulation of the stochastic series expansion technique for convenience. Namely, the following Hamiltonian
\begin{equation}\label{Eq1}
 \mathcal H=\sum_b^{N_b}J_b\bm{S_{i(b)}\cdot S_{j(b)}}
\end{equation}
describes spin models that include a number of bonds $N_b$ with site index $i(b)$ and spin operators \bm{$S_{i(b)}$} interacting with coupling strength $J_b$. For the present spin-ladder system with $N$ sites, we consider all spin-spin interactions between nearest neighbours to be antiferromagnetic ($J_b > 0$). Note that for the bonds connecting the spins along the legs of the spin ladder, we have $J_b = J$. The bonds located at the rungs of the system are randomly selected from a uniform bimodal distribution of the form
\begin{equation}\label{Eq2}
\mathcal P(J_b) = \frac{1}{2}[\delta(J_b-J_1)+\delta(J_b-J_2)],
\end{equation}
where $(J_1+J_2) / 2 = 1$ and $J_1 > J_2>0$ following previous practice \cite{Kanbur2,Malakis1,Vatansever1,Kanbur1}. The present study concerns only the case of $50\% / 50\%$ weak/strong bonds following the traditional bond disorder implementation commonly used in the case of classical Ising ferromagnets with disorder \cite{Fytas08}. With the help of the usual control parameter $r = (J_1 - J_2)/2$ that reflects the strength of the bond randomness on the rungs of the system we define $J_1 = 1+r \; (> 1)$ and $J_2 = 1-r \; (< 1)$. An example of the quenched bond disorder in the system is shown in Figure \ref{fig1}. For $r = 0$ the clean spin-ladder system is recovered, whereas the case $r = 1$ corresponds to the bond diluted case among the rungs of the system and is therefore beyond the scope of this study.

\subsection{Stochastic series expansion quantum Monte Carlo method}
\label{sec:method}

Within the framework of the stochastic series expansion  technique for the isotropic $S = 1/2$ Heisenberg antiferromagnetic model, it is possible that bond operators can be disentangled into diagonal ($H_{1,b}$) and off-diagonal ($H_{2,b}$) terms as follows
\begin{subequations}\label{Eq3}
\begin{equation}
H_{1,b}=\left(\frac14-S^z_{i(b)}S^z_{j(b)}\right) 
\end{equation}
\begin{equation}
H_{2,b}=\frac12\left(S^+_{i(b)}S^-_{j(b)}+S^-_{i(b)}S^+_{j(b)}\right).
\end{equation}
\end{subequations}
Benefiting from equation (\ref{Eq3}), we can thus rewrite Hamiltonian (\ref{Eq1}) as
\begin{equation}\label{Eq4}
 \mathcal H=-\sum_b^{N_b}J_b\left(H_{1,b}-H_{2,b}\right)+\mathrm{const.}
\end{equation}
We note here that the constant energy term is not necessary for the implementation of the algorithm (yet, it should be added when calculating the energy).

According to the stochastic series expansion quantum Monte Carlo protocol \cite{Sandvik1,Sandvik2,Sandvik3}, the partition function of the system can be Taylor expanded with a chosen spin basis as follows
\begin{equation}\label{Eq5}
  \mathcal Z=\sum_{\alpha,S_L}(-1)^{n_2}\beta^n\frac{(L-n)!}{L!}\left<\alpha\left|\prod_{p=0}^{L-1}J_{b(p)}H_{a(p),b(p)}\right|\alpha\right>.
\end{equation}
The summation appearing above refers to all configurations $\alpha$ and all possible operator strings $S_L$, including the unit operator $H_{0,0}$ to ensure that the length of the strings is fixed and also a unit bond coupling for convenient implementation; for this extra index, we have $J_0 = 1$. Here, $n$ is the number of non-unit operators in the string, while $n_2$ denotes the number of off-diagonal operators. Finally, as usual, $\beta$ is the reduced inverse temperature. Since the present system includes random bonds among the rungs, the non-zero weights are bond-dependent for an allowed spin configuration. Hence, the weight $W(\alpha, S_{L})$ can be expressed in the following manner
\begin{equation}\label{Eq6}
 W(\alpha, S_{L})=\left(\frac{\beta}{2}\right)^{n}\frac{(L-n)!}{L!}\prod_{p=0}^{L-1}J_{b(p)}.
\end{equation}

\section{Numerical simulations and observables}
\label{sec:numerical}

\subsection{Simulation details}
\label{sec:details}

Using the above scheme we generated numerical data for the quenched random-bond spin-ladder system for different system sizes at a wide temperature spectrum and for several values of the disorder strength parameter within the range $r =  \{0 - 0.9\}$. As already mentioned above, our geometrical setup consists of $L_x \times L_y$ systems, where $L_x$ defines the linear dimension along the legs of the lattice and is fixed at $L_x = 256$ during all simulations. We note in passing that we have performed some additional test simulations using $L_x$ values up to $512$ to identify possible finite-size and crossover effects in the observed quantities. However, our analysis, not shown here for brevity, disclosed that the results are almost $L_x$-independent so the considered $L_x = 256$ dimension is enough for an accurate determination of the properties. $L_y$ now stands for the number of legs of the ladder system. In fact, one of the most interesting aspects of the physics of quantum Heisenberg ladders is that their low-energy properties are distinct when comparing even and odd $L_y$ values \cite{White}. It is well-known that there is a spin gap for the case of even $L_y$ values, while ladder systems having odd $L_y$ values are gapless \cite{Sandvik1}. In this framework, we only studied systems with $L_y = \{2, 4, 6\}$. Closing, some numerical details: We performed simulations over $500$ independent random realizations for each pair of $(r, L_y)$ values. We applied boundary conditions which are periodic along the legs and free along the rungs of the spin ladder. In our protocol, the first $10^5$ Monte Carlo steps were discarded during the thermalization process and numerical data were collected and analyzed during the following $5\times 10^5$ Monte Carlo steps. We note in passing that the simulation time needed for a single realization on a node of a \emph{Dual Intel Xeon E5-2690 V4} processor was approximately $30$ minutes for the largest system size $(L_x = 256, L_y = 6)$ and the minimum temperature value that we have reached during the simulations. Finally, errors were estimated using the standard jackknife method \cite{Barkema,Binder}. 

\subsection{Observables}
\label{sec:observables}

The specific heat $C$ of the system is straightforward to estimate with the help of the number of 
non-unit operators $n$ in the operator sequence \cite{Sandvik1}
\begin{equation}\label{Eq7}
 C=\left<n^2\right>-\left<n\right>^2-\left<n\right>.
\end{equation}
Also, by constructing estimators from the Kubo integral, we gain access to static susceptibilities via \cite{Sandvik1}
\begin{equation}\label{Eq8}
\chi_{AB}=\int_0^\beta\,\mathrm{d}\tau\left<A(\tau)B(0)\right>.
\end{equation}
Here the integrand denotes the ensemble average of an imaginary time-dependent product with 
operators $A(\tau)=\mathrm{e}^{\tau H}\,A(0)\,\mathrm{e}^{-\tau H}$. For the case of diagonal operators $A$ and $B$ with eigenvalues $a(k)$ and $b(k)$, respectively, the static susceptibility is retrieved by including eigenvalues from all propagated states, as follows \cite{Sandvik4,Sandvik5}
\begin{eqnarray}
\label{Eq9}
 \chi_{AB}=\left\langle\frac\beta{n(n+1)} \left[ \left(\sum_{k=0}^{n-1}a(k)
 \right)\left(\sum_{k=0}^{n-1}b(k)
 \right)\right. \right. \nonumber\\
 \left. \left.  +\sum_{k=0}^{n-1}a(k)b(k)
 \right]
 \right\rangle.
\end{eqnarray}
For the total magnetization $M$, equation (\ref{Eq9}) reduces to 
the uniform susceptibility $\chi_u$ with $a(k)=b(k)=M$,
\begin{equation}\label{Eq10}
 \chi_u=\beta\left<M^2\right>.
\end{equation}
A low-temperature form of the uniform susceptibility of the spin-ladder system with an even number of rungs $L_y$ is given by \cite{Sandvik1}
\begin{equation}\label{Eq11}
    \chi_u = \frac{a}{\sqrt{T}}e^{-\Delta/T},
\end{equation}
where $\Delta$ is the well-known spin gap that can be extracted by linearization of equation (\ref{Eq11}) and making use of a least-squares fit -- typically and as it will be shown below, ground-state values of $\Delta$ are obtained while varying $(r, L_y)$ parameters. 

In addition to the physical observables introduced above, we also scrutinized in this work the thermal variations of the staggered susceptibility $\chi_s$ and structure factor $S$. For the staggered magnetization $M_s$, equation (\ref{Eq9}) with $a(k) = b(k)= M_s(k)$ provides the staggered susceptibility 
\begin{equation}
\chi_s(\pi,\pi)=\left<\frac\beta{n(n+1)}\left[\left(\sum_{k=0}^{n-1} M_s(k)
 \right)^2+\sum_{k=0}^{n-1} M_s^2(k)
 \right]
 \right>.
\end{equation}
Finally, the staggered structure factor can be computed via
\begin{equation}
 \label{eq:structure}
 S(\pi,\pi)=N\left< M_s^2\right>,
\end{equation}
where $N = L_x \times L_y$ the total number of spins on the system.

\begin{figure}%
\centering
\includegraphics[width=0.5\textwidth]{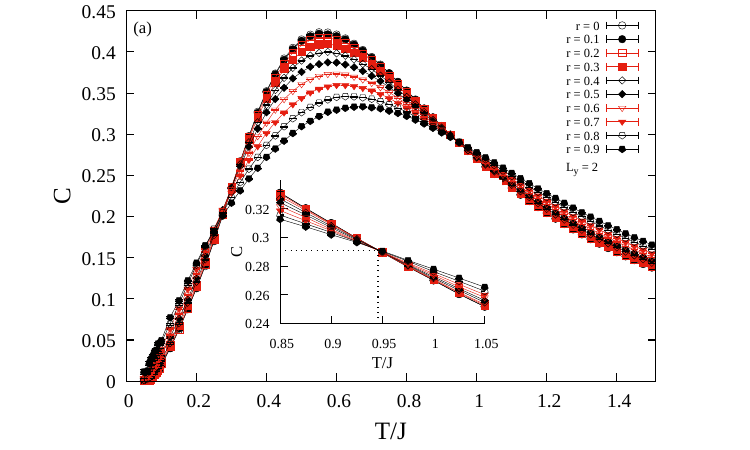}
\includegraphics[width=0.5\textwidth]{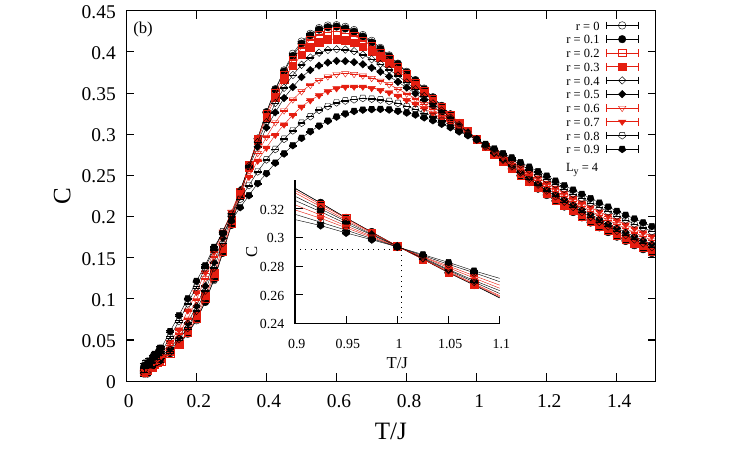}
\includegraphics[width=0.5\textwidth]{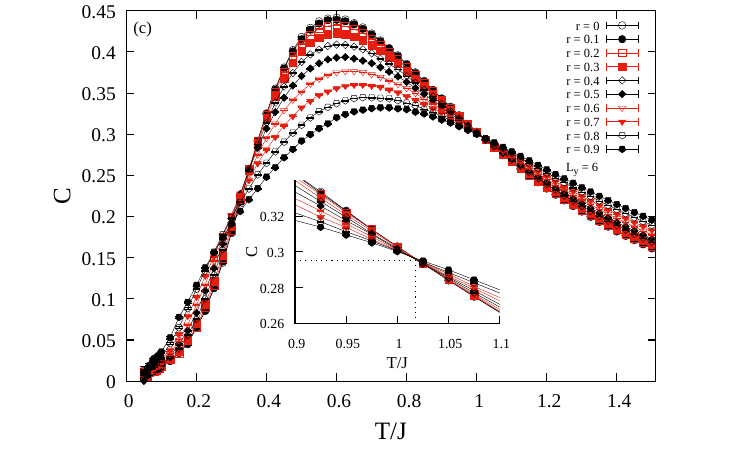}
\caption{Specific-heat curves vs. temperature for a wide range of the disorder parameter $r$ and three values of $L_y$. The insets are an enlargement of the intersection area (see also the discussion in the main text).}\label{fig2}
\end{figure}

\section{Results}
\label{sec:results}

We start the presentation of our results with Figure \ref{fig2}, showcasing the effects of disorder on the specific heat $C$ of the system for the three selected values of $L_y$ considered in this work. As it is evident from all panels, the specific-heat maxima decrease with increasing values of the disorder-strength ratio $r$ for all given values of $L_y$. In fact, the broad maxima of the curves shift towards a higher temperature regime and become flatter as $r$ gets larger. Interestingly, the considered system has a particular temperature point where the specific-heat curves intersect, taking the same value independent of $r$, but dependent on $L_y$. The inset in all three panels highlights the fine sweeping around these points and the vertical dashed lines mark their positions in the temperature axis, denoted as $(T/J)_{\rm cross}^{(C)}$. To obtain the crossing points for the specific-heat curves we used numerical data obtained for six $[r, r + 0.4]$ pairs of the disorder-strength ratio, where $r = \{0, 0.1, 0.2, 0.3, 0.4, 0.5\}$. We quote the estimates $(T/J)_{\rm cross}^{(C)} = 0.946(3)$, $1.004(1)$, and $1.018(2)$ for $L_y = 2$, $4$, and $L_y = 6$, respectively. Our analysis indicates that upon increasing $L_y$ the location of the crossing point shifts to higher temperatures. Note that the results obtained here for the case with $L_y = 2$ agree nicely within error bars with those from the analysis of Ref. \cite{Kanbur2}. 

\begin{figure}%
\centering
\includegraphics[width=0.5\textwidth]{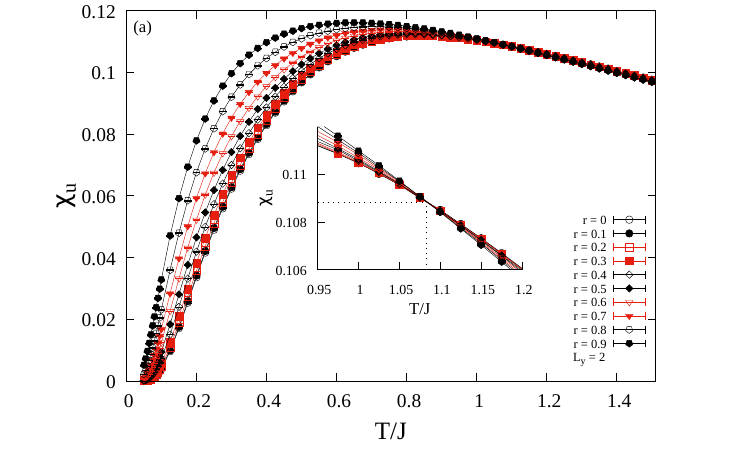}
\includegraphics[width=0.5\textwidth]{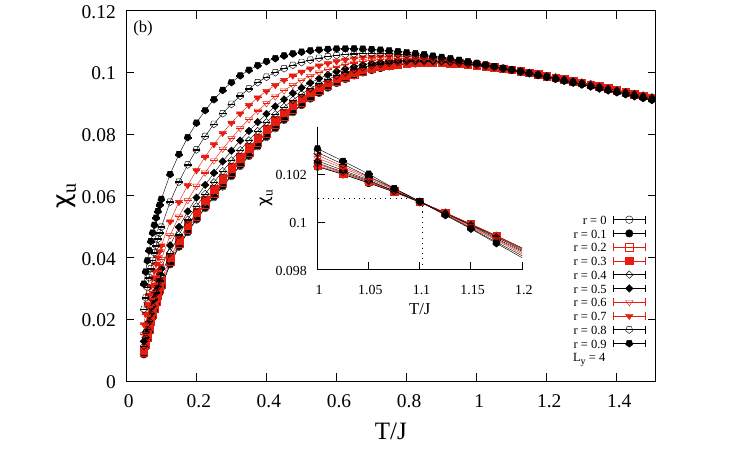}
\includegraphics[width=0.5\textwidth]{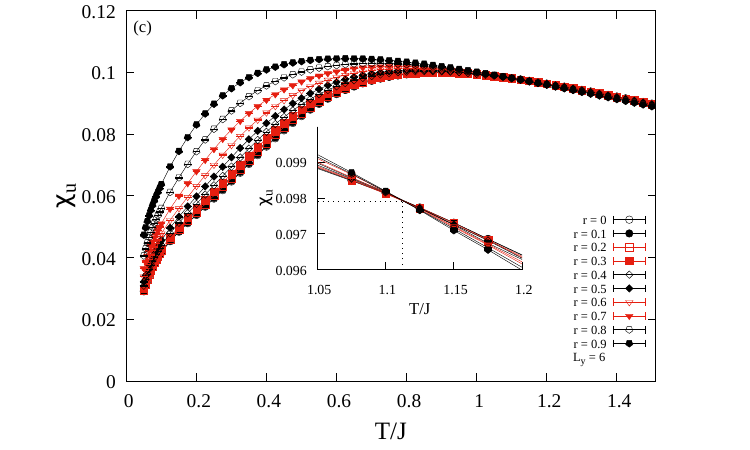}
\caption{Same as in Figure \ref{fig2}, but for the uniform susceptibility curves.}\label{fig3}
\end{figure}

We would like to underline that for this class of models including randomness on the local bonds, the presence of crossing points has been identified and studied in the context of experimental \cite{Vollhart,Chandra,Belevtsev,Brando} and theoretical \cite{Greger,Glocke,Georges} systems.  What is more, their physical origin has been scrutinized in detail for lattice models and continuum systems \cite{Vollhart}, and a wealth of numerical outcomes has been obtained in particular for the case of the half-filled Hubbard model at all dimensions \cite{Chandra}. Based on these studies one may safely conclude that the specific-heat values at these distinctive points are characterized by an almost universal value of $\sim 0.34/k_B$, whereas the corresponding temperatures are different at all dimensions. It is also worth stressing here that, for the present system, the rate of change in the specific-heat values as a function of the disorder parameter changes its sign at the intersection leading to a total entropy change of zero. Thereby, the crossing point of the specific heat may be also considered as an inflection point.

In analogy to Figure \ref{fig2}, we present in Figure \ref{fig3} the uniform susceptibility ($\chi_u$)-curves. Several comments are in order at this point: 

\begin{enumerate}
  \item At high temperatures, $\chi_u$ curves obtained for varying values of $r$ are almost identical and the typical Curie behaviour is present in the system for all values of $L_y$.
  \item In contrast to $C$, the broad maxima of $\chi_u$ shift to a lower temperature regime with increasing $r$, also for all values of $L_y$.
  \item Similarly to the observations recorded in Figure \ref{fig2}, there exist crossing points in the $\chi_u$ curves; see the corresponding insets of Figure \ref{fig3} for a zoom in of the criss-crossing area. To determine these points in $\chi_u$ we followed the procedure outlined above for the case of the specific heat. We quote the estimates $(T/J)_{\rm cross}^{(\chi_u)} = 1.083(3)$, $1.103(4)$, and $1.112(2)$ for $L_y = 2$, $4$, and $L_y = 6$, respectively.
  \item Even though the crossing temperatures obtained for the specific heat and uniform susceptibility (see Figures \ref{fig2} and \ref{fig3}) are slightly different from each other for a given $L_y$ value, the general shift trend to higher temperatures with increasing $L_y$ values is evident. We underline here that the absolute difference 
\[
\left|\left({T\over J}\right)_{\rm cross}^{(\chi_u)} - \left({T\over J}\right)^{(C)}_{\rm cross}\right| \longrightarrow 0,
\] 
as $L_{y}$ increases so that these discrepancies can be possibly attributed to the presence of finite-size effects.
\item Finally, an exponentially decreasing shift is noticeable at the relatively low-temperature regimes for all $L_y$ and $r$ values considered which clearly manifests the presence of a spin gap \cite{White}. 
\end{enumerate}

\begin{figure}%
\centering
\includegraphics[width=0.50\textwidth]{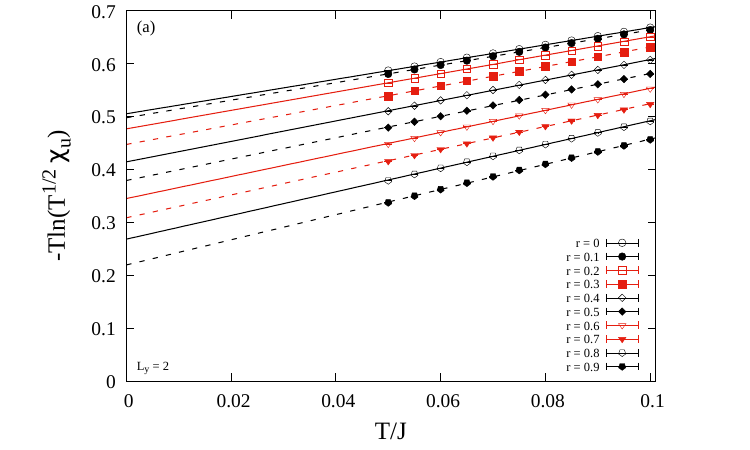}
\includegraphics[width=0.50\textwidth]{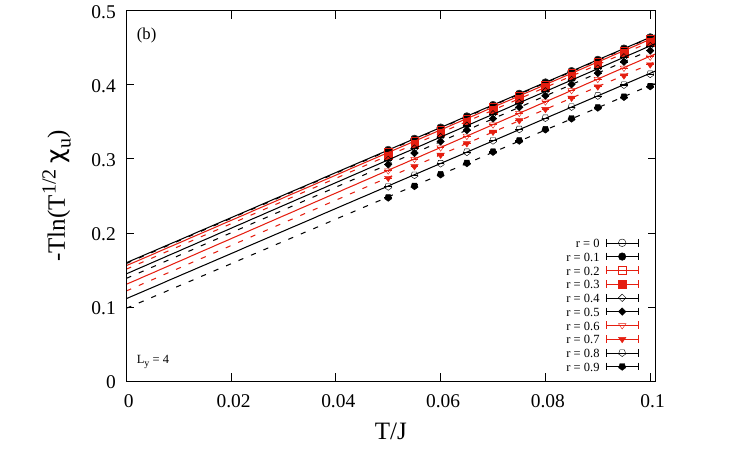}
\includegraphics[width=0.50\textwidth]{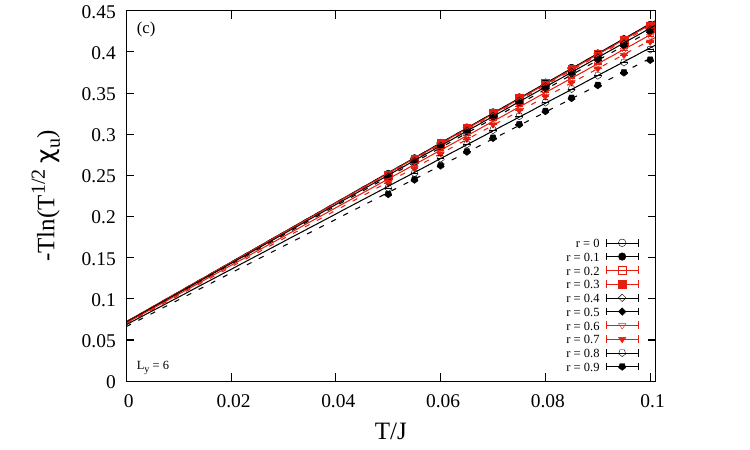}
\caption{Spin-gap fit lines of the form (\ref{Eq11}) for the clean case and all disorder ratios studied in this work for $L_y = 2$ (a), $L_y = 4$ (b), and $L_y = 6$ (c).}\label{fig4}
\end{figure}

\begin{figure}%
\centering
\includegraphics[width=0.50\textwidth]{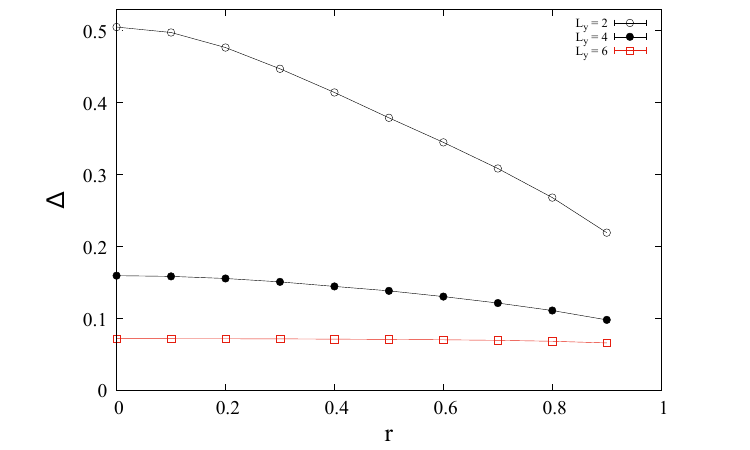}
\caption{Disorder-ratio dependence of the spin gap for all $L_y$ values considered. Lines are a simple guide to the eye, while error bars appear smaller than the symbol sizes used.}\label{fig5}
\end{figure}

Figure \ref{fig4} now displays the rescaled uniform susceptibility of the system, $-T\ln{(T^{1/2}\chi_{u}})$, for all studied values of $r$ and $L_y$ at the low-temperature regime, \emph{i.e.}, $(T/J) \leq 10^{-1}$. The solid and dashed lines are fits of the form (\ref{Eq11}) within the temperature range $5\times 10^{-2}\leq (T/J) \leq 10^{-1}$ where a clear linear response is observed. These extrapolations allow us to retrieve the spin-gap values $\Delta$ as $(T/J) \rightarrow 0$. The relevant $r$-dependence of the spin-gap values for varying values of $L_y$ is sketched in Figure \ref{fig5}. Remarkably, while inspecting Figure \ref{fig5} we deduce that when the amount of disorder among the rungs gets more robust, the spin-gap values tend to decrease, leading to an increment in the slopes of the relevant lines as shown in Figure \ref{fig4}. The most pronounced effect is found for the case $L_y = 2$ where the spin gap shows an almost linear dependence in $r$. Additionally, our simulation results suggest that upon increasing $L_y$ the impact of the disorder on the spin gap becomes less significant and, for the case with $L_y = 6$, almost irrelevant; see the red curve in Figure \ref{fig5} where $\Delta$ appears as almost constant. As it is well-known from previous studies \cite{frischmuth96,troyer94}, for a larger number of legs the drop in the uniform susceptibility sets in at smaller temperature values and is steeper. Hence, the spin gap is expected to decrease with increasing $L_y$ and the effect of randomness among the rungs of the system on the spin gap becomes trivial. We stress that our numerical estimates of $\Delta$ for the clean $(r = 0)$ cases are in good agreement also with previous works \cite{Sandvik1,Greven}.

\begin{figure}%
\centering
\includegraphics[width=0.50\textwidth]{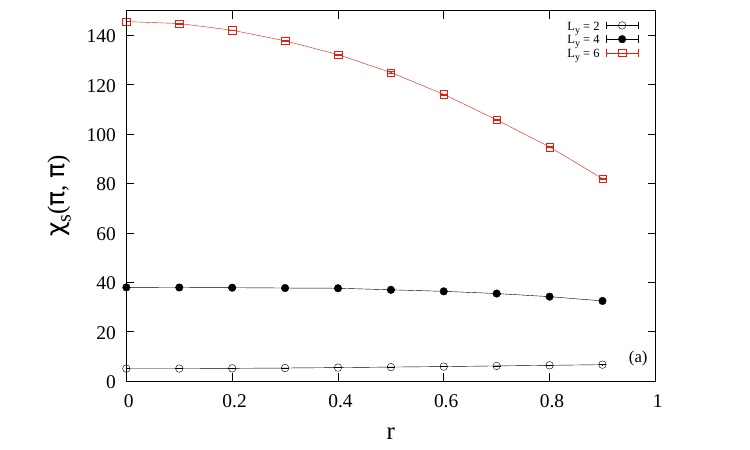}
\includegraphics[width=0.50\textwidth]{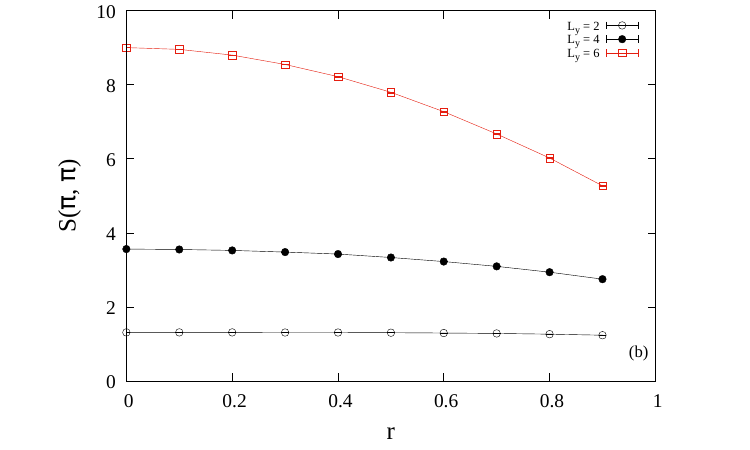}
\caption{Disorder-ratio dependence of the low-temperature behavior of the staggered susceptibility (a) and structure factor (b) for all $L_y$ values considered. Lines are a simple guide to the eye, while error bars appear smaller than the symbol sizes used.}\label{fig6}
\end{figure}

Along with the specific-heat and uniform susceptibility curves that show crossing points as detailed above, we have also considered in the last part of this section the staggered susceptibility ($\chi_s(\pi, \pi)$) and structure factor ($S(\pi, \pi)$) curves for the same set of system parameters. Our numerical results (not shown here for brevity) indicate that: (i) both quantities do not exhibit any instance of joining in the temperature plane for all studied values of the disorder-strength ratio $r$, and (ii) the structure-factor curves peak at a temperature well below that of the corresponding spin gap, an observation which is in full agreement with earlier studies \cite{Kanbur2,Greven}, giving further credit to the results presented in this work. In this framework, we consider it more beneficial to present in Figure \ref{fig6} the $r$-dependence of the low-temperature behaviour of $\chi_s(\pi, \pi)$, panel (a), and $S(\pi, \pi)$, panel (b), for $L_y = 2$, $4$, and $6$. Note that the numerical data were collected at the minimum temperature value reached during our simulations, \emph{i.e.}, $(T/J) = 5\times 10^{-2}$. The main aftermath from Figure \ref{fig6} is that while the $\chi_s(\pi, \pi)$ and $S(\pi, \pi)$ curves are almost flat for a wide range of $r$ values for the case $L_y = 2$, they tend to decrease with increasing $r$ for $L_y = 4$, and this trend becomes even more prominent for $L_y = 6$. Thus, in contrast to the case of the the spin-gap, see Figure \ref{fig5}, here the impact of disorder becomes much more striking with increasing $L_y$. To clarify this point it is possible to say that due to the presence of a spin gap in the system, the static structure factor is expected to remain finite, as $(T/J)\rightarrow 0$. With increasing $L_y$ for a chosen $r$ value, the spin gap tends to decrease, and as a result of this, $S(\pi, \pi)$ increases. Analogous conclusions have been reported for the case of spin-$1/2$ Heisenberg ladders yet under the presence of a different type of disorder \cite{Greven}.

\section{Summary}
\label{sec:summary}

To summarize, we investigated the effects of quenched bond randomness generated along the rungs of the antiferromagnetic spin-1/2 Heisenberg ladders using the stochastic series expansion quantum Monte Carlo technique. Numerical data were obtained for a wide spectrum of the disorder parameter $r$, from the clean case ($r = 0$) up to the bond-dilution limit ($r \rightarrow 1$) and for three values of $L_y$ of the system. Our simulations showed that while the specific-heat maxima decrease with increasing $r$ values for all studied cases of $L_y$, the uniform susceptibility curves appear to be almost identical, typical of the expected Curie behaviour. Only in the low-temperature regime, an increase in the $\chi_u$ values with increasing disorder was observed. Remarkably, for both $C$ and $\chi_u$ $r$-dependent curves the footprint of some characteristic crossing points in the temperature plane was unveiled, whose values are independent of $r$, however still moderately depend on $L_y$. Notably, the numerical values of the specific heat at these special points may be considered as (nearly) universal for the spin-ladder system under study. An analogous occurrence cannot be excluded also for the uniform susceptibility, although there appears to be a slim dependence of the $\chi_u$ values exactly at the crossing points on $L_y$, possibly attributed to finite-size effects. Another important result emerging in our study refers to the spin gap, whose presence was revealed from the exponentially decreasing current of the uniform susceptibility curves at the low-temperature regime. From our analysis we deduce that the spin gap decreases with increasing disorder parameter, analogously to the case of decreasing rung coupling values in the clean system, and that upon increasing $L_y$ the impact of disorder becomes less significant. Finally, we also studied the low-temperature behaviour of the staggered susceptibility and structure factor. In contrast to the spin-gap case, the impact of disorder in this case becomes more salient with increasing $L_y$. 

The results of this paper can be extended in several directions:

\begin{itemize}
  \item In terms of model systems, another fruitful platform would be the spin-$1$ quantum Heisenberg ladder, where the presence of the reported crossing points (and possible universality) has not been yet clarified.
  \item In terms of the diffused randomness, it would be beneficial to consider also other types of disorder distributions, such as bond dilution or non-magnetic impurity atoms which could mimic some aspects of quantum magnetic materials.
  \item Finally, the role of an additional external magnetic field on the spin ladder could provide insight to other physical phenomena, not touched upon the current work.
\end{itemize}

\begin{acknowledgements}
The numerical calculations reported in this paper were performed at T\"{U}B\.{I}TAK ULAKBIM (Turkish agency), High Performance and Grid Computing Center (TRUBA Resources). The work of GGG is  supported by the Bulgarian National Science Fund (grant KP-06N42-2 is acknowledged). The work of NGF was supported by the  Engineering and Physical Sciences Research Council (grant EP/X026116/1 is acknowledged).
\end{acknowledgements}

\section*{\large{Declarations}}

\textbf{Author contributions} All authors contributed equally to the paper.

\textbf{Data availability statement} This manuscript has no associated data or the data will not be deposited. [Authors’ comment: This is a theoretical study with no experimental data. All numerical data generated during this study are available on reasonable request.].

\textbf{Conflict of interest} The authors have no conflicts to disclose.

\end{document}